\documentclass[a4paper,12pt]{article}
\usepackage{epsfig}
\usepackage{citesort}
\date{\today}
 \normalsize

\def\be{\begin{equation}}
\def\ee{\end{equation}}
\def\bea{\begin{eqnarray}}
\def\eea{\end{eqnarray}}

\def\sss{\scriptscriptstyle}
\def\nn{\nonumber}
\def\bra#1{\left\langle #1\right|}
\def\ket#1{\left| #1\right\rangle}
\newcommand{\bers}{\begin{eqnarray*}}
\newcommand{\eers}{\end{eqnarray*}}
\newcommand{\bt}{\begin{itemize}}
\newcommand{\et}{\end{itemize}}
\def\bvv{B \to V_1 V_2}
\def\lr{(\delta^d_{LR})_{23}}
\def\rl{(\delta^d_{RL})_{23}}
\def\th1{\theta_{LR}}
\def\th2{\theta_{RL}}
\def\lsim{\raise0.3ex\hbox{$\;<$\kern-0.75em\raise-1.1ex\hbox{$\sim\;$}}}
\def\gsim{\raise0.3ex\hbox{$\;>$\kern-0.75em\raise-1.1ex\hbox{$\sim\;$}}}

\def\ie{{\it i.e.}}

\def\epjc#1#2#3{{ Eur.\ Phys.\ J.}\ {\bf C#1}, #3, (#2)}

\def\npb#1#2#3{{ Nucl.\ Phys.} {\bf B#1}, #3 (#2)}
\def\plb#1#2#3{{ Phys.\ Lett.} {\bf #1B}, #3 (#2)}
\def\prd#1#2#3{{ Phys.\ Rev.} {\bf D#1}, #3 (#2)}
\def\newprd#1#2#3{{ Phys.\ Rev.} {\bf D#1}, #3 (#2)}
\def\prl#1#2#3{{ Phys.\ Rev.\ Lett.} {\bf #1}, #3 (#2)}

\def\zpc#1#2#3{{ Zeit.\ Phys.} {\bf C#1}, #3 (#2)}
\begin{document}
%\begin{titlepage}
\renewcommand{\thefootnote}{\fnsymbol{footnote}}
\vspace{.3cm} {\large \begin{center} {\bf Supersymmetry and CP
violation in $B^0_s - \bar{B}^0_s$ mixing and $B^0_s \to J/\psi
\phi$ decay}
\end{center}}
\vspace{.3cm}
\begin{center}
Alakabha Datta$^1$ and Shaaban Khalil$^{2,3}$\\[5mm]
$^1$\emph{Dept of Physics and Astronomy, 108 Lewis Hall,
University of Mississippi, Oxford, MS 38677-1848, USA.}\\
$^2$\emph{Center for Theoretical Physics at the
British University in Egypt, Sherouk City, Cairo 11837, Egypt.\\
$^3$ Department of Mathematics, Ain Shams University, Faculty of
Science,
Cairo, 11566, Egypt.}
\end{center}
\vspace{.5cm}
\abstract{Supersymmetric contributions to time independent
asymmetry in $B^0_s \to J/\psi \phi$ process are analyzed in the
view of the recent Tevatron experimental measurements. We show
that the experimental limits of the mass difference $\Delta
M_{B_s}$ and the mercury EDM significantly constrain the SUSY
contribution to $B^0_s -\bar{B}^0_s$ mixing, so that $\sin 2
\beta_s \lsim 0.1$. We also point out that the one loop SUSY
contribution to $B^0_s \to J/\psi \phi$ decay can be important and
can lead to large indirect CP asymmetries which are different for
different polarization states. These new physics effects in the
decay amplitude can be consistent with CP measurements in the
$B_d$ system.
% Therefore, it can account for a large value of $\sin
%2\beta_s$, consistent with the recent experimental results by $D0$
%and $CDF$.
}
%\end{titlepage}
%%%%%%%%%%%%%%%%%%%%%%%%%%%%%%%%%%%%%%%%%%%%%%%%%%%%%%
%
\section{\large{\bf Introduction}}%
Recently, CDF and $D0$ collaborations have announced the
observation of CP violation in $B_s^0 -\bar{B}_s^0$ mixing. The
following results,
for $B^0_s$-mixing CP violating phase, have been reported \cite{Aaltonen:2007he,:2008fj}:%
\bea %
2 \beta_s &=& 0.57~^{+0.30}_{-0.24}~ ({\rm stat.})~
^{+0.02}_{-0.07}~({\rm syst.})~~~~~~~~~~~~~~~~~~~~~~ (DO), \\
2 \beta_s &\in& \left[0.32 , 2.82\right] (68\%)
~~~~~~~~~~~~~~~~~~~~~~~~~~~~~~~~~~~~
(CDF).%
\eea %
These results indicate that the phase $\beta_s$ deviates more that
$3\sigma$ from the Standard Model (SM) prediction
\cite{Bona:2008jn}. Therefore, the experimental observation of CP
violation in $B^0_s$ mixing, along with the Belle and Baber
measurement for direct and indirect CP asymmetries of $B_d$
decays, open the possibility of probing new physics effect at low
energy.

It is a common feature for any physics beyond the SM to possess
additional sources of CP violation besides the SM phase in quark
mixing matrix. In supersymmetric extension of the SM, the soft
SUSY breaking terms are in general complex and can give new
contributions to CP violating processes. The SUSY CP violating
phases can be classified as flavor independent phases, like the
phases of the gaugino masses and $\mu$ term, and flavor-dependent
phases, like the phases of the off-diagonal $A$-terms. The flavor
independent phases are stringently constrained by the experimental
limits on electric dipole moment (EDM) of electron and neutron.
However, the flavor dependent phases are much less constrained.
This may imply that SUSY CP violation has a flavor off diagonal
character just as in the Standard Model. In this case the origin
of CP violation is closely related to the origin of the flavor
structures rather than the origin of SUSY breaking
\cite{Abel:2001vy}.

The SUSY flavor dependent phases can induce sizeable contributions
to direct and indirect CP asymmetries of $B_d$ decays
\cite{Khalil:2002fm,Khalil:2003bi,Khalil:2005qg}, as in $B_d \to
\phi K_S$, $B_d \to \eta' K_S$ and $B_d \to K \pi$ which show some
discrepancy with the SM expectation. In this paper we revisit the
supersymmetric contributions to $B_s^0 -\bar{B}_s^0$ mixing. We
investigate the possibility that SUSY may be responsible for the
large observed value of $B_s$ mixing phase without enhancing the
mass difference $\Delta M_s$ over the measured value. In addition,
we analyze the one loop SUSY contribution to $B^0_s \to J/\psi
\phi$ decay, which turns out to be important and can lead to a
large indirect CP asymmetry.

The paper is organized as follows. In section 2 we analyze the
possible new physics contributions to $B_s^0 -\bar{B}_s^0$ mixing
and indirect CP asymmetry of $B_s^0 \to J/\psi \phi$, taking into
account the constraints imposed by the experimental measurments of
the mass difference $\Delta M_{B_s}$ and the mercury EDM. In
section 3 we discuss the supersymmetric contributions to effective
Hamiltonian for $\Delta B = 2$ and $\Delta B = 1$ transitions. In
section 4 we show that the mercury EDM impose stringent
constraints on the supersymmetric contribution to the phase
$\beta_s$, such that the $B_s^0$ mixing phase can not exceed 0.1.
In section 5 we analyze the supersymmetric contribution to the
$B_s^0 \to J/\psi \phi$ decay. We emphasize that the one loop SUSY
contribution to $B_s^0 \to J/\psi \phi$ can be important and lead
to large indirect CP asymmetries which are in general different
for different polarization states. Finally, we give our
conclusions in section 6.

%%%%%%%%%%%%%%%%%%%%%%%%%%%%%%%%%%%%%%%%%%%%%%%%%%%%%%%%%%%%%%
%%%%%%%%%%%%%%%%%%%%%%%%%%%%%%%%%%%%%%%%%%%%%%%%%%%%%%%%%%%%%%
%
\section{\large{\bf $B_s^0 -\bar{B}_s^0$ mixing and CP asymmetry in $B_s^0 \to J/\psi \phi$}}%
In the the $B_s^0$ and $\bar{B}_s^0$ system, the flavor
eigenstates are given by $B^0_s=(\bar{b} s)$ and
$\bar{B}_s^0=(b\bar{s})$. The corresponding mass eigenstates are
defined as $B_L=p B_s^0 - q \bar{B}_s^0$ and $B_H =p B_s^0 + q
\bar{B}_s^0$, where $L$ and $H$ refer to light and heavy mass
eigenstates respectively. The mixing angles $q$ and $p$ are
defined in terms of the transition matrix element ${\cal M}_{12}=
\langle B_s^0 \vert H_{eff}^{\Delta B=2} \vert \bar{B}_s^0
\rangle$, where $H_{eff}^{\Delta B=2}$ is the effective
Hamiltonian responsible for $\Delta B=2$ transitions:%
\be%
\frac{q}{p}= \sqrt{\frac{{\cal M}_{12}^*}{{\cal M}_{12}}},%
\ee%
where we assumed that $\Delta \Gamma_{B_s} \ll \Delta M_{B_s}$ and
$\Delta \Gamma_{B_s} \ll \Gamma_{B_s}^{total}$. The strength of
$B_s^0 - \bar{B}_s^0$
mixing is described by the mass deference%
\be%
\Delta M_{B_s}= M_{B_H} - M_{B_L}= 2 {\rm Re}\left[ \frac{q}{p}
{\cal M}_{12}\right]= 2 \vert {\cal M}_{12}(B_s)\vert .%
\ee%

The decay $B^0_s \to J/\psi \phi$ involves vector-vector final
states with three polarization amplitudes. Therefore, an angular
distribution is necessary to separate out the three polarizations
for a measurement of indirect CP violation without dilution. The
amplitudes for the decay of $B_s^0 \to f $ and $\bar{B}_s^0 \to f$
are given by $A^{\lambda}(f) =\langle f \vert H_{eff}^{\Delta B=1}
\vert B_s^0 \rangle$ and $\bar{A}^{\lambda}(f) =\langle f \vert
H_{eff}^{\Delta B=1} \vert \bar{B}_s^0 \rangle$ with %
\be%
\bar{\rho}^{\lambda}(f) = \frac{\bar{A}^{\lambda}(f)}{A^{\lambda}(f)} = \frac{1}{\rho^{\lambda}(f)}. %
\ee%
Here, $\lambda$, is the polarization index.
Therefore, the source of CP violation in decays to CP eigenstates
with oscillation are: oscillation if $q/p \neq 1$, decay if
$\bar{\rho}^{\lambda}(f) \neq 1$, both oscillation and decay if
$\{q/p,\bar{\rho}^{\lambda}(f)\} \neq 1$. The time dependent CP asymmetry of
$B_s^0 \to J/\psi \phi$, for each polarization state $\lambda$, is given by %
\bea %
A^{\lambda}_{J/\psi \phi} (t) &=& \frac{\Gamma^{\lambda}(\bar{B}_s^0(t) \to J/\psi
\phi) - \Gamma^{\lambda}(B_s^0(t) \to J/\psi \phi)}{\Gamma^{\lambda}(\bar{B}_s^0(t)
\to J/\psi \phi) + \Gamma^{\lambda}(B_s^0(t) \to J/\psi \phi)}, \nonumber\\
&=&C^{\lambda}_{J/\psi \phi} \cos \Delta M_{B_s} t + S^{\lambda}_{J/\psi \phi} \sin \Delta M_{B_s} t ,%
\eea%
where $C^{\lambda}_{J/\psi \phi}$ and $S^{\lambda}_{J/\psi \phi}$ represent the direct
and the mixing CP asymmetry, respectively and they are given by%
\be%
C^{\lambda}_{J/\psi \phi} = \frac{\vert \bar{\rho}^{\lambda}(J/\psi \phi)\vert^2
-1}{\vert \bar{\rho}^{\lambda}(J/\psi \phi)\vert^2 +1},~~~~~~~~~~~ S^{\lambda}_{J/\psi
\phi} = \eta^{\lambda}\frac{2 {\rm Im} \left[\frac{q}{p} \bar{\rho}^{\lambda}(J/\psi
\phi)\right]}{\vert \bar{\rho}^{\lambda}(J/\psi \phi)\vert^2 +1},%
\ee%
where $\eta^{\lambda}$ is $\pm$ depending on the polarization
states. In the SM, the mixing CP asymmetry in $B^0_s \to J/\psi
\phi$ process is the same for all polarization, to a very good
approximation, up to a sign. Hence we will omit the polarization
index when discussing the SM results. We have in the SM,
\be%
\sin 2 \beta_s =S_{J/\psi \phi}.%
\ee%
If $\rho(J/\psi \phi) =1$, which is the case in SM, then $\beta_s$ is defined as $ 2
\beta_s =\arg \left[{\cal M}_{12}(B_s)\right]$.

In the SM, the mass difference is given by %
\be %
\Delta M_{B_s}^{SM} = \frac{G_F^2}{6 \pi^2} \eta_B m_{B}
(\hat{B}_{B_s} F_{B_s}^2) M_W^2 \vert V_{ts} \vert^2 S_0(x_t). %
\ee %
One may estimate the SM contribution to $\Delta M_{B_s}$ through
the ratio $\Delta M_{B_s}^{SM}/\Delta M_{B_d}^{SM}$, where the
uncertainties due to short-distance effect cancel:
\be%
\frac{\Delta M_{B_s}^{SM}}{\Delta M_{B_d}^{SM}} =
\frac{M_{B_s}}{M_{B_d}} \frac{B_{B_s} f_{B_s}^2}{B_{B_s}
f_{B_s}^2} \frac{\vert V_{ts}\vert^2}{\vert V_{td}\vert^2}. %
\ee%
We can assume that $\Delta M_{B_d}^{SM}=  \Delta M_{B_d}^{exp}
\simeq 0.507 {\rm ps}^{-1}$. Thus, for quark mixing angle $\gamma
\simeq 67^{~\circ}$, one finds $\Delta M_{B_s}^{SM} \simeq 15 {\rm
ps}^{-1}$, which is consistent with the recent results reported by
CDF and $D0$:
\bea %
\Delta M_{B_s} &=& 17.77\pm 0.10 ({\rm stat.}) \pm 0.07({\rm syst.})~~~~~~~~~~~~~~~~~~ (CDF), \\
\Delta M_{B_s} &=& 18.53\pm 0.93 ({\rm stat.}) \pm 0.30({\rm
syst.})~~~~~~~~~~~~~~~~~~ (CDF).
\eea %

On the other hand, the SM contribution ($\rho(J/\psi \phi)=1$) to
the
CP asymmetry $S_{J/\psi \phi}$ is given by %
\be %
S_{J/\psi \phi} = \sin 2 \beta_{s}^{SM}, ~~~~  {\rm with} ~~~~
\beta_s^{SM} = \arg\left(\frac{-V_{cs} V_{cb}^*}{V_{ts}
V_{tb}^*}\right)
\simeq {\cal O}(0.01), %
\ee %
where $V_{ij}$ are the elements of the CKM quark mixing matrix.
This result clearly conflicts with the experimental measurements
reported in Eqs.(1,2). Therefore, a confirmation for these
measurements would be no doubt signal for new physics beyond the
SM. As indicated above, $S_{J/\psi \phi}$ carries a polarization
index corresponding to the three final state polarization, however
in the SM the mixing induced asymmetries are the same( up to a
sign) for the three polarizations.

\begin{figure}[t]
\begin{center}
\epsfig{file=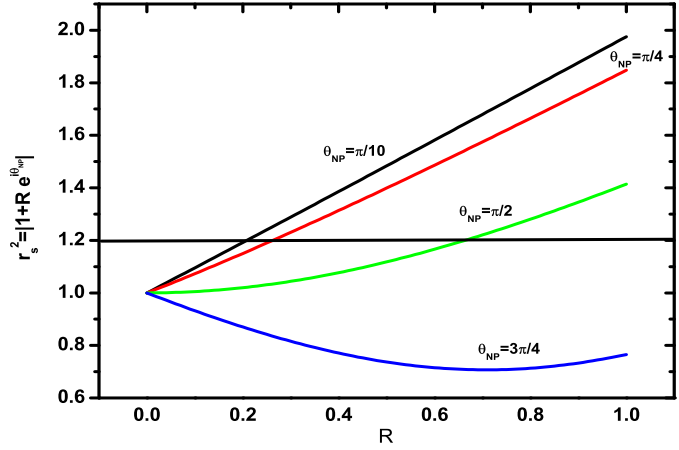, width=10cm,height=7cm,angle=0}
\end{center}
\vspace{-1cm} \caption{The constraint on $R = \vert A_{NP}/A_{SM}
\vert$ in case of $\theta= \pi/10, \pi/4, \pi/2$ and $3\pi/4$.}
\label{fig:R}
\end{figure}

In a model independent way, the effect of new physics (NP), with
$\rho(J/\psi \phi)=1$, can be described by the dimensionless
parameter $r_s^2$ and a phase $2
\theta_s$ defined as follows: %
\be %
r_s^2 e^{2 i \theta_s} = \frac{{\cal M}_{12}(B_s)}{{\cal
M}^{SM}_{12}(B_s)} = 1+ \frac{{\cal M}^{NP}_{12}(B_s)}{{\cal
M}^{SM}_{12}(B_s)} , %
\ee %
Therefore, $\Delta M_{B_s} = 2 \vert {\cal M}^{SM}_{12}(B_s)\vert
r_s^2 = \Delta M^{SM}_{B_s} r_s^2$. In this respect, $r_s^2$ is
bounded by $r_s^2 \lsim \Delta M^{exp}_{B_s}/\Delta M^{SM}_{B_s}
\lsim 1.2$. This constrains the ratio between the NP
 and SM amplitudes defined as, $R=\vert A_{NP}/A_{SM} \vert$, as follow: %
\be %
\left \vert 1+ R e^{i \theta_{NP}} \right \vert \lsim 1.2 %
\ee%
Note that for vanishing NP phase, \ie $\theta_{NP}=0$, one find
that $R \lsim 0.2$. However, for $\theta_{NP} \neq 0$, the
constrain on $R$ is relaxed as shown in Fig. 1. It is clear that
$R$ can be of order one if the NP phase is tuned to be within the
range: $\pi/2 < \theta_{NP} < \pi$, .

In the presence of NP contribution, the CP asymmetry $B^0_s \to
J/\psi \phi$ is modified and now we have %
\be %
S_{J/\psi \phi} = \sin 2 \beta_{eff} = \sin(2 \beta_s^{SM} + 2
\theta_s), %
\ee %
where %
\be %
2 \theta_s = \arg \left( 1+ R e^{\theta_{NP}} \right). %
\ee %
Therefore, in order to enhance the NP effects, large values of $R$
are required. Now we consider the effect of NP that leads to ${\rm
Im}[\rho(J/\psi \phi)] \neq 1$.  Let us write the amplitude as %
\be%
\bar{A}^{\lambda}(J/\psi \phi) = \bar{A}^{\lambda}_{SM}(J/\psi \phi) +
\bar{A}^{\lambda}_{NP}(J/\psi \phi),%
\ee%
and define, %
\be %
\frac{A^{\lambda}(J/\psi \phi)}{A^{\lambda}_{SM}(J/\psi \phi)} =
S_A^{\lambda} e^{i \theta^{\lambda}_A},%
\label{amp_ratio}
\ee %
where $\theta^{\lambda}_A$ is a weak phase, $\lambda$ is the polarization index,
and we have assumed that the strong phases in the amplitude ratio cancel.
One can now write $\bar{\rho}(J/\psi \phi)$ as %
\be%
\bar{\rho}(J/\psi \phi) = e^{-2 i \theta^{\lambda}_A}.
%\frac{V_{cb}
%V_{cs}^*}{V_{cb}^* V_{cs}}, %
\ee%
Thus, one obtains,%
\be%
\frac{q}{p} \bar{\rho}(J/\psi \phi) = e^{-2i
(\beta_{SM} + \theta_s + \theta^{\lambda}_A)}. %
\ee%
In this case, the CP asymmetry $B^0_s \to J/\psi \phi$ is modified
and now we have,
\be %
S_{J/\psi \phi} =  \sin(2 \beta_s^{SM} + 2
\theta_s + 2 \theta^{\lambda}_A). %
\ee %
However, as pointed out in Ref.\cite{Khalil:2002fm}, this
parametrization is true only when the strong phase of the full
amplitude is assumed to be the same as the SM amplitude. In fact,
as discussed in Ref.~{\cite{datta}}, the NP strong phase can be
different and is generally smaller than the SM strong phase thus
invalidating the assumption about strong phases made in
Eq.(\ref{amp_ratio}). In general, the SM and NP
amplitude can be parameterized as:%

\be%
A^{\lambda}_{{\rm SM}}= \vert A^{\lambda}_{{\rm SM}} \vert e^{i \delta^{\lambda}_{{\rm SM}}},
~~~~~ A^{\lambda}_{{\rm NP}}= \sum_i \vert A^{\lambda}_{{i\rm NP}} \vert e^{i\theta^{\lambda}_{{i\rm
NP}}} e^{i \delta^{\lambda}_{{i \rm NP}}}, %
\ee %
where $\delta^{\lambda}_{i\rm NP}$ are the strong phases and
$\theta^{\lambda}_{{i \rm NP}}$ are the CP violating phase. If
there is one dominant NP amplitude then we can parametrize the NP
amplitude as \bea A^{\lambda}_{{\rm NP}}&= &  \vert
A^{\lambda}_{{\rm NP}} \vert e^{i\theta^{\lambda}_{{i\rm NP}}}e^{i
\delta^{\lambda}_{{i \rm NP}}} \ \label{NPamp}. \eea Thus, the CP
asymmetry $S_{J/\psi \phi}$ can
be approximately written as: %
\be %
S_{J/\psi \phi}^{\lambda} = \sin(2 \beta_s^{SM} + 2 \theta_s) + 2 r_A^{\lambda} \cos(2 \beta_s^{SM}
+ 2 \theta_s) \sin \theta^{\lambda}_{{\rm NP}} \cos \delta^{\lambda},
\label{newbetas} %
\ee%
where $r_A^{\lambda}= \vert A_{{\rm NP}}^{\lambda}/A_{{\rm
SM}}^{\lambda} \vert$ and $\delta^{\lambda}=
\delta^{\lambda}_{{\rm SM}}-\delta^{\lambda}_{{\rm NP}}$. Here
$\lambda$ represents the various polarization states of the
vector-vector final state.

In the SUSY case, considered in this paper, there will be two
dominant operators. In this case we can write the new physics
amplitude as, \be A^{\lambda}_{{\rm NP}}=  \vert
A^{\lambda}_{{1\rm NP}} \vert e^{i\theta^{\lambda}_{{1\rm NP}}}
e^{i \delta^{\lambda}_{{1 \rm NP}}} +\vert A^{\lambda}_{{2\rm NP}}
\vert e^{i\theta^{\lambda}_{{2\rm NP}}} e^{i \delta^{\lambda}_{{2
\rm NP}}} \ \label{susynp}.%
\ee%
Now using the result in Ref.~{\cite{datta}}, we will neglect the
NP strong phase and hence the new physics amplitude can be
rewritten as an effective single NP amplitude, \bea
A^{\lambda}_{{\rm NP}} &= & \vert A^{\lambda}_{{\rm NP}} \vert e^{i\theta^{\lambda}_{{\rm NP}}} \nonumber\\
\tan{\theta^{\lambda}_{{\rm NP}}} & = & \frac{
\vert A^{\lambda}_{{1\rm NP}}\vert \sin{\theta^{\lambda}_{1 \rm NP}}+
\vert A^{\lambda}_{{2\rm NP}}\vert \sin{\theta^{\lambda}_{2 \rm NP}}
}
{
 \vert A^{\lambda}_{{1\rm NP}}\vert \cos{\theta^{\lambda}_{1 \rm NP}}
+\vert A^{\lambda}_{{2\rm NP}}\vert\cos{\theta^{\lambda}_{2 \rm NP}}
}
\nonumber\\
\vert A^{\lambda}_{{\rm NP}}\vert & = & \sqrt{ \left( \vert
A^{\lambda}_{{1\rm NP}}\vert\sin{\theta^{\lambda}_{1 \rm NP}}+
\vert A^{\lambda}_{{2\rm NP}}\vert\sin{\theta^{\lambda}_{2 \rm
NP}} \right)^2+ \left( \vert A^{\lambda}_{{1\rm
NP}}\vert\cos{\theta^{\lambda}_{1 \rm NP}}+ \vert
A^{\lambda}_{{2\rm NP}}\vert\cos{\theta^{\lambda}_{2 \rm
NP}}\right)^2 }~~~~~~ \label{effamp} \eea
 Hence the expression in Eq.(\ref{newbetas}) can still be used
provided we set the NP strong phases to zero.
%
%%%%%%%%%%%%%%%%%%%%%%%%%%%%%%%%%%%%%%%%%%%%%%%%%%%%%%
\section{\large{\bf Supersymmetric contributions to $\Delta B=2$ and $\Delta B=1$ transitions}}%
In this section, we analyze the SUSY contribution to the $B_s^0
-\bar{B}_s^0$ mixing and $B_s^0 \to J/\psi \phi$ decay. As pointed
out in Ref.\cite{Ball:2003se}, gluino exchanges through $\Delta
B=2$ box diagrams give the dominant contribution to $B_s^0
-\bar{B}_s^0$ mixing, while the chargino exchanges are subdominant
and can be neglected. The general $H_{\mathrm{eff}}^{\Delta B=2}$
induced by gluino
exchanges can be expressed as %
\be H_{\mathrm{eff}}^{\Delta B=2} =
\sum_{i=1}^5 C_i(\mu) Q_{i}(\mu) + \sum_{i=1}^3 \tilde{C}_i(\mu)
\tilde{Q}_i(\mu) + h.c. , %
\ee %
where $C_i(\mu)$, $\tilde{C}_i(\mu)$, $Q_i(\mu)$ and
$\tilde{Q}_i(\mu)$ are the Wilson coefficients and operators
respectively normalized at the scale $\mu$, with, %
\bea%
Q_1 &=& \bar{s}^{\alpha}_L \gamma_{\mu} b_L^{\alpha}~
\bar{s}^{\beta}_L
\gamma_{\mu} b_L^{\beta},\\
Q_2 &=& \bar{s}^{\alpha}_R b_L^{\alpha}~ \bar{s}^{\beta}_R
b_L^{\beta},\\
Q_3 &=& \bar{s}^{\alpha}_R b_L^{\beta}~ \bar{s}^{\beta}_R
b_L^{\alpha},\\
Q_4 &=& \bar{s}^{\alpha}_R b_L^{\alpha}~ \bar{s}^{\beta}_L
b_R^{\beta},\\
Q_5 &=& \bar{s}^{\alpha}_R b_L^{\beta}~ \bar{s}^{\beta}_L
b_R^{\alpha}. \eea %
In addition, the operators $\tilde{Q}_{1,2,3}$ are obtained from
$Q_{1,2,3}$ by exchanging $L \leftrightarrow R$. The results for
the gluino contributions to the above Wilson coefficients at SUSY
scale, in the frame work of the mass
insertion approximation, are give by \cite{masiero}%
\bea %
C_1^{\tilde{g}}\!&=&\!-\frac{\alpha_s^2}{216 m_{\tilde{q}}^2}
\left[ 24 x f_6(x) + 66 \tilde{f}_6(x) \right] (\delta_{23}^d)^2_{LL},~~~~ \\
C_2^{\tilde{g}}\! &=&\!-\frac{\alpha_s^2}{216 m_{\tilde{q}}^2}
204 x f_6(x) (\delta_{23}^d)^2_{RL}, \\
C_3^{\tilde{g}}\! &=&\!-\frac{\alpha_s^2}{216 m_{\tilde{q}}^2} 36
x f_6(x)
(\delta_{23}^d)^2_{RL} , \\
C_4^{\tilde{g}}\! &=&\!-\frac{\alpha_s^2}{216 m_{\tilde{q}}^2}
\left\{ \left[ 504 x f_6(x) -72
 \tilde{f}_6(x) \right] (\delta_{23}^d)_{LL}(\delta_{23}^d)_{RR} -132
\tilde{f}_6(x) (\delta^d_{23})_{LR} (\delta^d_{23})_{RL} \right\},~~~~ \\
C_5^{\tilde{g}}\! &=&\!-\frac{\alpha_s^2}{216 m_{\tilde{q}}^2}
\left\{ \left[ 24 x f_6(x) +120 \tilde{f}_6(x) \right]
(\delta_{23}^d)_{LL} (\delta_{23}^d)_{RR} -180 \tilde{f}_6(x)
(\delta^d_{23})_{LR} (\delta^d_{23})_{RL} \right\}. %
\eea %
where $x=m^2_{\tilde{g}}/m_{\tilde{q}}^2$ with $m_{\tilde{g}}$ and
$m_{\tilde{q}}$ are the gluino mass and the average squark mass,
respectively. The expressions for the functions $f_6(x)$ and
$\tilde{f}_6(x)$ can be found in Ref.\cite{masiero}. The Wilson
coefficients $\tilde{C}_{1,2,3}$ are obtained by interchanging the
$L\leftrightarrow R$ in the mass insertions appearing in
$C_{1,2,3}$.

Note that the mass insertions $(\delta^d_{23})_{LL}
(\delta^d_{23})_{RR}$ may give the dominant contribution to the
transition matrix element, due to its large coefficient in
$C_4^{\tilde{g}}$. In order to connect $C_i(M_S)$ at the SUSY
scale $M_S$ with the corresponding low energy ones $C_i(\mu)$ with
$\mu \sim \mathcal{O}(m_b)$, one has to solve the RGE for the
Wilson coefficients. Also the matrix elements of the operators
$Q_i$ can be found in Ref.\cite{Becirevic:2001jj}.

Now, we turn to supersymmetric contribution to the amplitude for $B_s \to J/\psi
\phi$. It turns out that the gluino exchanges through $\Delta B=1$
penguin diagrams gives the dominant contributions to this process.
The effective Hamiltonian for the $\Delta B=1$ transitions through the
penguin process can, in general, be expressed as,
\begin{equation}
\mathcal{H}^{\Delta B=1}_{\mbox{eff}}= \sum_{i=3}^{6}C_iO_i+C_gO_g
\sum_{i=3}^{6}\tilde{C}_i\tilde{O}_i+\tilde{C}_g\tilde{O}_g ,
\end{equation}
where
\begin{eqnarray}
O_3 &=&\bar{s}_L^{\alpha}\gamma^{\mu} b_L^{\alpha}
    \bar{c}_L^{\beta}\gamma_{\mu} c_L^{\beta}, \\
O_4 &=&\bar{s}^{\alpha}_L\gamma^{\mu} b^{\beta}_L
    \bar{c}^{\beta}_L\gamma_{\mu} c^{\alpha}_L, \\
O_5 &=&\bar{s}^{\alpha}_L\gamma^{\mu} b^{\alpha}_L
    \bar{c}^{\beta}_R\gamma_{\mu} c^{\beta}_R, \\
O_6 &=&\bar{s}^{\alpha}_L\gamma^{\mu} b^{\beta}_L
    \bar{c}^{\beta}_R \gamma_{\mu} c^{\alpha}_R, \\
O_{g} &=& \frac{g_s}{8\pi^2}m_b\bar{s}^{\alpha}_L \sigma^{\mu\nu}
\frac{\lambda^A_{\alpha\beta}}{2}b^{\beta}_R
G^A_{\mu\nu}\label{og}.
\end{eqnarray}
At the first order in the mass insertion approximation, the gluino
contributions to the Wilson coefficients $C_{i,g}$ at the SUSY
scale $M_S$ are given by \cite{masiero}
\begin{eqnarray}
C_3 (M_S) &=& \frac{\alpha_s^2}{m_{\tilde{q}}^2}
(\delta_{LL}^d)_{23} \left[\frac{1}{9} B_1(x) + \frac{5}{9} B_2(x)
+
\frac{1}{18} P_1(x) +\frac{1}{2} P_2(x) \right],\nonumber\\
C_4 (M_S) &=& \frac{\alpha_s^2}{m_{\tilde{q}}^2}
(\delta_{LL}^d)_{23} \left[ \frac{7}{3} B_1(x) - \frac{1}{3}
B_2(x) - \frac{1}{6} P_1(x)
-\frac{3}{2} P_2(x) \right],\nonumber\\
C_5 (M_S) &=& \frac{\alpha_s^2}{m_{\tilde{q}}^2}
(\delta_{LL}^d)_{23} \left[ -\frac{10}{9} B_1(x) - \frac{1}{18}
B_2(x) + \frac{1}{18} P_1(x)
+\frac{1}{2} P_2(x) \right], \label{eq:0047}\\
C_6 (M_S) &=& \frac{\alpha_s^2}{m_{\tilde{q}}^2}
(\delta_{LL}^d)_{23} \left[ \frac{2}{3} B_1(x) - \frac{7}{6}
B_2(x) - \frac{1}{6} P_1(x)
-\frac{3}{2} P_2(x) \right],\nonumber\\
C_g (M_S)\!\! &=&\!\!\frac{\alpha_s \pi} {m_{\tilde{q}}^2}\!\left[
\!(\delta_{LL}^d)_{23}\left( \frac{1}{3} M_3(x)\! + \!3
M_4(x)\right)\!+\! (\delta_{LR}^d)_{23}\frac{m_{\tilde{g}}}{m_b}
\left(\!\frac{1}{3} M_3(x) \!+\! 3
M_2(x)\right)\!\right]\!,\nonumber
\end{eqnarray}

The absolute values of the mass insertions $(\delta_{AB}^d)_{23}$,
with $A,B=(L,R)$ are constrained by the experimental results for
the branching ratio of the  $B \to X_s \gamma$ decay. These
constraints are very weak on the $LL$ and $RR$ mass insertions and
the only limits we have come from their definition, $\vert
(\delta_{LL,RR}^d)_{23} \vert < 1$. The $LR$ and $RL$ mass
insertions are more constrained and, for instance with
$m_{\tilde{g}}\simeq m_{\tilde{q}}\simeq 500$ GeV, one obtains
$\vert(\delta_{LR,RL}^d)_{23} \vert \lsim 1.6 \times 10^{-2}$
\cite{masiero,Khalil:2005qg}. Note that, although, the $LR(RL)$
mass insertion are constrained severely their effects to the decay
are enhanced by a large factor $m_{\tilde{g}}/m_b$ as can be seen
from the above expression for $C_{g}(M_S)$.

In this respect, the phase of $(\delta_{LR}^d)_{23}$,
$(\delta_{LL}^d)_{23}$ and $(\delta_{RR}^d)_{23}$ are the relevant
CP violating phases for our process. In the next section, we
discuss possible constraints imposed on these phases by the
mercury EDM.

\section{\large{\bf Mercury EDM versus large $ B^0_s-\bar{B}^0_s$ mixing phase}}%

It has been pointed out \cite{Hisano:2004tf,Abel:2004te} that
large values of $(\delta^d_{23})_{RR}$ may enhance the
chromo-electric dipole moment of the strange quark which is
constrained by the experimental bound on the EDM of mercury atom
$H_g$. In this section we show that the $H_g$ EDM imposes a
constraint on ${\rm Im}[(\delta_{LL}^d)_{23}
(\delta_{RR}^d)_{23}]$, which may limit the supersymmetric
contribution to the $B^0_s-\bar{B}^0_s$ mixing.

In the chiral lagrangian approach, the mercury EDM is given by
\cite{Hisano:2004tf}%
\be%
d_{Hg} = -e \left(d^C_d -d^C_u - 0.012 d^C_s\right) \times 3.2
\times
10^{-2}. %
\ee%
The chromelectric EDM of the strange quark $d^C_s$ is given by%
\be %
d^C_s = \frac{g_s}{\alpha_s}{4\pi}
\frac{m_{\tilde{g}}}{m_{\tilde{d}^2}} {\rm Im}
(\delta^d_{22})_{LR} M_2(x)~, %
\ee %
where $x=m_{\tilde{g}}^2/m_{\tilde{d}}^2$. For $m_{\tilde{d}}=500$
GeV and $x=1$, the experimental limit on $H_g$ EDM leads to the
following constraint on $(\delta^d_{23})_{LR}$:
\be %
{\rm Im}(\delta^d_{22})_{LR} < 5.6 \times 10^{-6}~. %
\ee%
The mass insertion $(\delta^d_{22})_{LR}$ may be generated
effectively through three mass insertions as follows: %
\be%
(\delta^d_{22})_{LR}\simeq (\delta^d_{23})_{LL}
(\delta^d_{33})_{LR} (\delta^d_{32})_{RR},%
\ee%
where $(\delta^d_{33})_{LR}  \simeq  \frac{m_b(A_b - \mu \tan
\beta)}{m_{\tilde{d}}^2} \simeq {\cal O}(10^{-2})$. Therefore, the
$H_g$ EDM imposes the following constraint on the $LL$ and $RR$
mixing between the second and the third generations:%
\be %
{\rm Im}\left[(\delta^d_{23})_{LL}
(\delta^d_{23})^{\dag}_{RR}\right]
\lsim 5.6 \times 10^{-4} .%
\ee%
If one assumes that $(\delta^d_{23})_{LL} \sim \lambda^2$  with
negligible weak phase, then he gets the following bound on
the $(\delta^d_{23})_{RR}$ mass insertion: %
\be %
\vert (\delta^d_{23})_{RR} \vert ~\sin \left(\arg
[(\delta^d_{23})_{RR}]\right)
\lsim 10^{-2} . %
\ee %
Therefore, in case $\vert (\delta^d_{23})_{RR}\vert \sim {\cal
O}(0.01)$, the associated weak phase is essentially unconstrained.
However, if $\vert (\delta^d_{23})_{RR}\vert \sim {\cal O}(0.1)$,
the the weak phase is constrained to be of order $0.1$. In both
cases, this will limit the SUSY contributions to the $B^0_s -
\bar{B}_s^0$ mixing phase.

We start our analysis for SUSY contribution to $\sin 2 \beta_s$ by
assuming that $B^0_s -\bar{B}^0_s$ mixing may receive a
significant SUSY contribution, while the decay of $B^0_s \to
J/\psi \phi$ is dominated by the SM. Therefore, we have ${\rm
Im}[\rho(J/\psi \phi)]=0$ and the induced CP asymmetry is given by
$S_{J/\psi \phi} = \sin(2 \beta^{{\rm SM}}_s + 2 \theta_s)$. As an
example for the SUSY contribution, we consider $m_{\tilde{q}}=500$
GeV and $x=1$, which leads to the following expression for
$R=\vert {\cal
M}_{12}^{SUSY}/{\cal M}_{12}^{SM} \vert$ \cite{Ball:2003se}: %
\bea %
R &=& \left \vert 1.44 \left[ (\delta^d_{23})_{LL}^2 +
(\delta^d_{23})_{RR}^2 \right] + 27.57 \left[
(\delta^d_{23})_{LR}^2 + (\delta^d_{23})_{RL}^2 \right] - 44.76
\left[(\delta^d_{23})_{LR}
(\delta^d_{23})_{RL}\right] \right.\nonumber\\
&-& \left.175.79\left[(\delta^d_{23})_{LL}
(\delta^d_{23})_{RR}\right] \right \vert.
\eea %
From this equation, it is noticeable that the dominant
contribution to the $B_s^0-\bar{B}_s^0$ mixing is due to the mass
insertions $(\delta^d_{23})_{LL} (\delta^d_{23})_{RR}$.

If one assumes that $(\delta^d_{23})_{LL}$ is induced by the
running from the high scale, where left-handed squark masses are
universal, down to the electrweak scale, then one finds
$(\delta^d_{23})_{LL} \sim \lambda^2 \sim 0.04$. With a small
source of non-universality in the right-handed squark sector, one
can easily get $(\delta^d_{23})_{RR}$ of order ${\cal O}(0.1)$.
Therefore, one gets $R \sim 0.7$. However in this case, the $H_g$
EDM implies that: ${\rm arg}[(\delta^d_{RR})_{23}] \lsim 0.1$,
which limits significantly the SUSY effect for enhancing $\sin 2
\beta_s$.

\begin{figure}[t]
\begin{center}
\epsfig{file=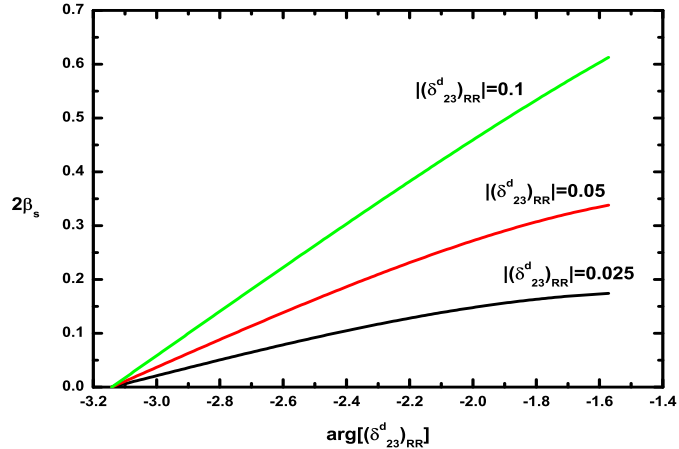, width=10cm,height=7cm,angle=0}
\end{center}
\vspace{-1cm} \caption{The $B^0_s - \bar{B}^0_s$ mixing phase as
function of the $\arg [(\delta^d_{23})_{RR}]$ (in radians) for
$\vert (\delta^d_{23})_{RR}\vert = 0.025, 0.05$ and $0.1$.}
\label{fig:betas}
\end{figure}

In Fig. \ref{fig:betas}, we present our results for the $B^0_s -
\bar{B}^0_s$ mixing phase $2 \beta_s$ as a function of
$\arg[(\delta^d_{23})_{RR}]$ for $\vert (\delta^d_{23})_{RR}\vert
= 0.025, 0.05$ and $0.1$. At these values the ratio $R$ is of
order $\lsim 0.17, 0.35$ and $0.7$ respectively. As can be seen
from this figure, the values of $B^0_s$ mixing phase, which are
consistent with the Hg EDM constraints, are typically of order
$\lsim 0.1$.
Therefore, one concludes that the SUSY contribution through the
$B^0_s - \bar{B}^0_s$ mixing  implies limited enhancement for
$\sin 2 \beta_s$ and thus cannot account for the new experimental
results reported in Eq.(1,2). Moreover, a salient feature of this
scenario with large $RR$ mixing is that it predicts a reachable
mercury EDM in the future experiments.

\section{\large{\bf SUSY contribution to $\bar{B}^0_s \to J/\psi \phi$ decay}}%

In this section we will consider SUSY contribution to the decay
$\bar{B}^0_s \to J/\psi \phi$. However, let us discuss the
complexities in analyzing new physics effects in the decay
amplitude for vector-vector final state\cite{sinha}.

Consider a $\bvv$ decay which is dominated by a single weak decay
amplitude within the SM. This holds for processes which are
described by the quark-level decays ${\bar b} \to {\bar c} c {\bar
s}$ which is the underlying quark transition in $\bar{B}^0_s \to
J/\psi \phi$. In this case, the weak phase of the SM amplitude is
zero in the standard parametrization~\cite{pdg}. Suppose now that
there is a single dominant new physics amplitude, with a different
weak phase, that contributes to the decay. This indeed will be the
case for the SUSY contribution to $\bar{B}^0_s \to J/\psi \phi$.
The decay amplitude for each of the three possible helicity states
may be written as
\bea A_\lambda \equiv {\rm Amp} (\bvv)_\lambda &=& a_\lambda e^{i
\delta_\lambda^a} + b_\lambda e^{i\phi} e^{i \delta_\lambda^b} ~,
\nn\\
{\bar A}_\lambda \equiv {\rm Amp} ({\bar B} \to V_1 V_2)_\lambda
&=& a_\lambda e^{i \delta_\lambda^a} + b_\lambda e^{-i\phi} e^{i
\delta_\lambda^b} ~, \label{amps} \eea
where $a_\lambda$ and $b_\lambda$ represent the SM and NP amplitudes,
respectively, $\phi$ is the new-physics weak phase, the
$\delta_\lambda^{a,b}$ are the strong phases, and the helicity index
$\lambda$ takes the values $\left\{ 0,\|,\perp \right\}$. Using CPT
invariance, the full decay amplitudes can be written as
\bea {\cal A} &=& {\rm Amp} (\bvv) = A_0 g_0 + A_\| g_\| + i \,
A_\perp
g_\perp~, \nn\\
{\bar{\cal A}} &=& {\rm Amp} ({\bar B} \to V_1 V_2) = {\bar A}_0
g_0 + {\bar A}_\| g_\| - i \, {\bar A}_\perp g_\perp~,
\label{fullamps} \eea
where the $g_\lambda$ are the coefficients of the helicity
amplitudes written in the linear polarization basis. The
$g_\lambda$ depend only on the angles describing the kinematics
\cite{glambda}. Eqs.~(\ref{amps}) and (\ref{fullamps}) above
enable us to write the time-dependent decay rates as
\be \Gamma(\bar{B}^0_s(t) \to V_1V_2) = e^{-\Gamma t}
\sum_{\lambda\leq\sigma} \Bigl(\Lambda_{\lambda\sigma} \pm
\Sigma_{\lambda\sigma}\cos(\Delta M t) \mp
\rho_{\lambda\sigma}\sin(\Delta M t)\Bigr) g_\lambda g_\sigma ~.
\label{decayrates} \ee
Thus, by performing a time-dependent angular analysis of the decay
$B(t) \to V_1V_2$, one can measure 18 observables. These are:
\bea
\Lambda_{\lambda\lambda}=\displaystyle
\frac{1}{2}(|A_\lambda|^2+|{\bar A}_\lambda|^2),~~&&
\Sigma_{\lambda\lambda}=\displaystyle
\frac{1}{2}(|A_\lambda|^2-|{\bar A}_\lambda|^2),\nn \\[1.ex]
\Lambda_{\perp i}= -\!{\rm Im}({ A}_\perp { A}_i^* \!-\! {\bar
A}_\perp {{\bar A}_i}^* ),
&&\Lambda_{\| 0}= {\rm Re}(A_\| A_0^*\! +\! {\bar A}_\| {{\bar A}_0}^*
), \nn \\[1.ex]
\Sigma_{\perp i}= -\!{\rm Im}(A_\perp A_i^*\! +\! {\bar A}_\perp
{{\bar A}_i}^* ),
&&\Sigma_{\| 0}= {\rm Re}(A_\| A_0^*\!-\! {\bar A}_\| {{\bar A}_0}^*
),\nn\\[1.ex]
\rho_{\perp i}\!=\! {\rm Re}\!\Bigl(\frac{q}{p} \!\bigl[A_\perp^*
{\bar A}_i\! +\! A_i^* {\bar A}_\perp\bigr]\Bigr),
&&\rho_{\perp \perp}\!=\! {\rm Im}\Bigl(\frac{q}{p}\, A_\perp^*
{\bar A}_\perp\Bigr),\nn\\[1.ex]
\rho_{\| 0}\!=\! -{\rm Im}\!\Bigl(\frac{q}{p}[A_\|^* {\bar A}_0\! +
\!A_0^* {\bar A}_\| ]\Bigr),
&&\rho_{ii}\!=\! -{\rm Im}\!\Bigl(\frac{q}{p} A_i^* {\bar A}_i\Bigr),
  \label{eq:obs}
\eea
where $i=\{0,\|\}$. In the above, $q/p$ is the weak phase factor
associated with $B^0_s$-${\bar B}^0_s$ mixing. For $B^0_s$ meson,
$q/p = \exp({-2\,i\beta_s})$. Note that $\beta_s$ may include NP
effects in $B^0_s$-$\bar{B}^0_s$ mixing. Note also that the signs
of the various $\rho_{\lambda\lambda}$ terms depend on the
CP-parity of the various helicity states. We have chosen the sign
of $\rho_{00}$ and $\rho_{\|\|}$ to be $-1$, which corresponds to
the final state $ J/\psi \phi$.

Not all of the 18 observables are independent. There are a total of
six amplitudes describing $\bvv$ and ${\bar B} \to V_1 V_2$ decays
[Eq.~(\ref{amps})]. Thus, at best one can measure the magnitudes and
relative phases of these six amplitudes, giving 11 independent
measurements.

The 18 observables given above can be written in terms of 13
theoretical parameters: three $a_\lambda$'s, three $b_\lambda$'s,
$\beta_s$, $\phi$, and five strong phase differences defined by
$\delta_\lambda \equiv \delta_\lambda^b - \delta_\lambda^a$, $\Delta_i
\equiv \delta_\perp^a - \delta_i^a$. The explicit expressions for the
observables can be found in Ref~\cite{sinha}.
In the presence of new physics, one
cannot extract the phase $\beta_s$. There are 11 independent
observables, but 13 theoretical parameters. Since the number of
measurements is fewer than the number of parameters, one cannot
express any of the theoretical unknowns purely in terms of
observables. In particular, it is impossible to extract $\beta_s$
cleanly.

In the absence of NP, the $b_\lambda$ are zero in
Eq.~(\ref{amps}). The number of parameters is then reduced from 13
to 6: three $a_\lambda$'s, two strong phase differences
($\Delta_i$), and $\beta_s$. It is straightforward to show that
all six parameters can be determined cleanly in terms of the
observables. This is exactly what is done in the experimental
measurements to measure $\beta_s$, the value of which appears to
be inconsistent with the SM. This might indicate new non SM phase
in $B_s$ mixing or NP in the decay amplitude in which case the
general angular analysis in Eq.~(\ref{decayrates}) should be used.
In the presence of NP, the indirect CP asymmetries for the various
polarization states are not longer the same as it is in the SM( up
to a sign).

\begin{figure}[t]
\begin{center}
\epsfig{file=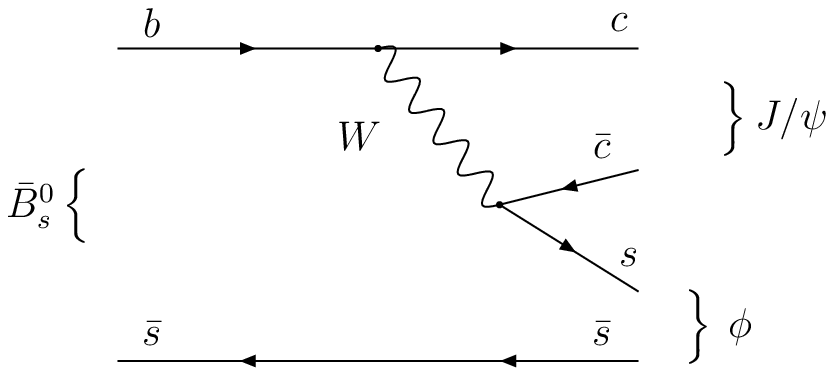, width=7cm,height=4cm,angle=0}~~~~
\epsfig{file=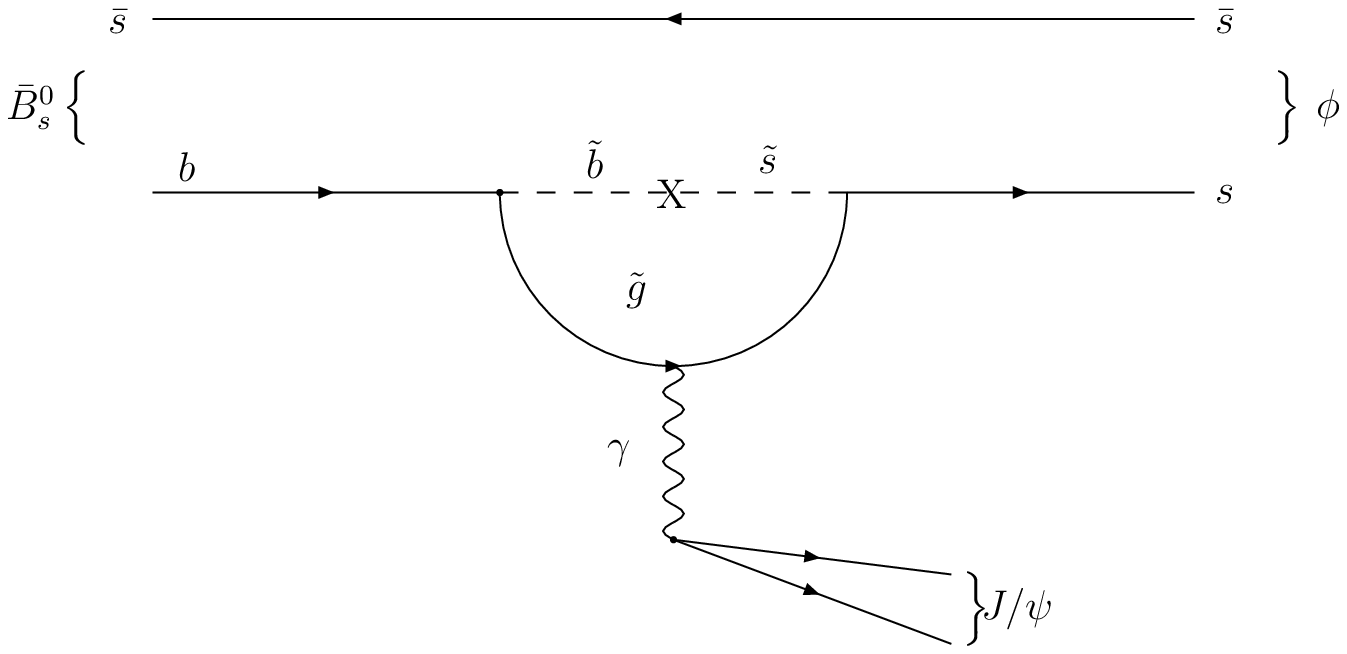, width=7cm,height=4cm,angle=0 }
\end{center}
\label{fig.3} \vspace{-0.5cm} \caption{SM tree level (left) and
SUSY one loop (right) contributions to $\bar{B}^0_s \to J/\psi
\phi$ decay.} \label{fig:betas}
\end{figure}

In this section we will consider the scenario where SUSY gives
significant contribution to both $B^0_s -\bar{B}^0_s$ mixing and
the decay of $B^0_s \to J/\psi \phi$.  In this case, the induced
CP asymmetry is given by Eq.(\ref{newbetas}). As shown in
Fig.(\ref{fig.3}), the SM the decay of $B^0_s \to J/\psi \phi$
takes place at tree level through the $b \to c$ transition. While
the dominant SUSY contribution to this decay is given by the one
loop level of gluino exchange for $b\to s$ transition. It is
interesting to note that the SM amplitude is proportional to $G_F
\times V_{bc} V_{cs} \sim 10^{-7}$, while the SUSY amplitude is
given in terms of $\alpha_s^2/m_{\tilde{q}}^2
\left((\delta^d_{LR})_{23} \times m_{\tilde{g}}/m_b\right)$.
Therefore, although SUSY contribution is a loop level, it can be
important relative to
 the SM one. In this respect, it is important to
consider the impact of this contribution on the induced CP
asymmetry $S^{\lambda}_{J/\psi \phi}$, as the phase of the mass insertion
$(\delta^d_{LR})_{23}$ is not constrained by EDM.

Let us now write down the SM and SUSY contribution to $B^0_s(p)
\to J/\psi(k_1, \epsilon_1) \phi(k_2,\epsilon_2)$, where we have
labelled the momentum and polarization of the final state
particles. To proceed with our calculation, we will first specify
the momentum and polarization vectors of the final-state
particles. We will work in the rest frame of the $B^0_s$ meson.
We define the momentum and polarization of the vector $\phi$ meson
as \cite{VT}
\begin{eqnarray}
k_2^{\mu} & = & (E_{\phi}, 0, 0, -k) \nonumber\\
\varepsilon_2^{\mu}(0) & = & \frac{1}{m_{\phi}}(-k,0, 0, E_{\phi}) \nonumber\\
\varepsilon_2^{\mu}(\mp) &= &\frac{1}{\sqrt{2}}(0, \mp 1, -i, 0) ~,\
\label{pol_phi}
\end{eqnarray}
The momentum and polarization vectors for $J/\psi$ are defined as,
\begin{eqnarray}
k_1^{\mu} & = & (E_{J/\psi}, 0, 0, k) \nonumber\\
\varepsilon_1^{\mu}(0) & = & \frac{1}{m_{J/\psi}}(k,0, 0,
E_{J/\psi}) ~,\nonumber\\
\varepsilon_1^{\mu}(\pm) &= &\frac{1}{\sqrt{2}}(0, \mp 1, -i, 0) ~,\
\label{pol_psi}
\end{eqnarray}
The general amplitude for $\bar{B}^0_s(p) \to J/\psi(k_1,
\varepsilon_1) \phi(k_2,\varepsilon_2)$, can be expressed
as\cite{TP}, \be \bar{A} = \bar{a} \, \varepsilon_1^* \cdot
\varepsilon_2^* + {\bar{b} \over m_{B_s}^2} (p\cdot
\varepsilon_1^*) (p\cdot \varepsilon_2^*) + i {\bar{c} \over
m_{B_s}^2} \epsilon_{\mu\nu\rho\sigma} p^\mu q^\nu
\varepsilon_1^{*\rho} \varepsilon_2^{*\sigma} ~, \label{abcdefs}
\ee where $q=k_1-k_2$. {}For angular analysis it is useful to use
the linear polarization basis. In this basis, one decomposes the
decay amplitude into components in which the polarizations of the
final-state vector mesons are either longitudinal ($A_0$), or
transverse to their directions of motion and parallel ($A_\|$) or
perpendicular ($A_\perp$) to one another. One writes
\cite{DDLR,CW}
\be \bar{A} = \bar{A}_0 \varepsilon_1^{*\sss L} \cdot
\varepsilon_2^{*\sss L} - {1 \over \sqrt{2}} \bar{A}_\|
{\vec\varepsilon}_1^{*\sss T} \cdot
  {\vec\varepsilon}_2^{*\sss T}
- {i \over \sqrt{2}} \bar{A}_\perp {\vec\varepsilon}_1^{*\sss T}
\times
  {\vec\varepsilon}_2^{*\sss T} \cdot {\hat p} ~,
\ee
where ${\hat p}$ is the unit vector along the direction of motion
of $V_2$ in the rest frame of $V_1$, $\varepsilon_i^{*\sss L} =
{\vec\varepsilon}_i^* \cdot {\hat p}$, and
${\vec\varepsilon}_i^{*\sss T} = {\vec\varepsilon}_i^* -
\varepsilon_i^{*\sss L} {\hat p}$. $\bar{A}_0$, $\bar{A}_\|$,
$\bar{A}_\perp$ are related to $a$, $b$ and $c$ of
Eq.~(\ref{abcdefs}) via
\be \bar{A}_\| = \sqrt{2} \bar{a} ~,~~~ \bar{A}_0 = -\bar{a} x -
{m_1 m_2 \over m_{\sss B}^2} \bar{b} (x^2 - 1) ~,~~~ \bar{A}_\perp
= 2\sqrt{2} \, {m_1 m_2 \over m_{\sss B}^2} \bar{c} \sqrt{x^2 - 1}
~, \label{Aidefs} \ee
where $x = k_1 \cdot k_2 / (m_1 m_2)$. (A popular alternative basis is
to express the decay amplitude in terms of helicity amplitudes
$A_\lambda$, where $\lambda = 1, 0 ,-1$ \cite{KP,DDLR}. The helicity
amplitudes can be written in terms of the linear polarization
amplitudes via $A_{\pm 1} = (A_\| \pm A_\perp)/\sqrt{2}$, with $A_0$
the same in both bases.)

We will now proceed to calculate the polarization dependent CP
asymmetry given in Eq.~\ref{newbetas}. We will use factorization
to calculate the ratio $r_A^{\lambda}= \vert A_{{\rm
NP}}^{\lambda}/A_{{\rm SM}}^{\lambda} \vert$. In factorization
there are no strong phases and we will keep them as a free unknown
parameter in the expression for $S_{J/\psi \phi}^{\lambda}$ in
Eq.~(\ref{newbetas}). The amplitude for the process $\bar{B}_s(p)
\to J/\psi(k_1, \varepsilon_1) \phi(k_2,\varepsilon_2)$, in the
SM, is given by,
\be \bar{A}[\bar{B}_s \to J/\psi \phi] = \frac{G_F}{\sqrt{2}} X
L_{J/\psi}, \ee
with
\bea
X & = & V_{cb} V_{cs}^* a_2 - \sum_{q=u,c,t} V_{qb} V_{qs}^* (a_3^q +
a_5^q + a_7^q + a_9^q) ~, \nonumber\\
L_{J/\psi} & = & m_{J/\psi} g_{J/\psi} \varepsilon_1^{*\mu}
\bra{\phi} \bar{s} \gamma_{\mu} ( 1-\gamma_5 ) b \ket{\bar{B}_s},\
\label{sm} ~. \eea where $a_2=c_2+\frac{c_{1}}{N_c}$ and for $i >
2$, $a_i=c_i+ \frac{c_{i+i}}{N_c}$, with $c_i$ being the Wilson's
coefficient. Here $g_{J/\psi}$ is the $J/\psi$ decay constant
defined in the usual manner.

We can simplify $X$ using several facts. First $a_2$ is much
larger than $a_i^t$ with $i=3,5,7,9$ \cite{BuraseffH}. Second, in
the penguin contributions in Eq.~(\ref{sm}) we have included the
rescattering contributions from the tree operators. However these
are small and the contributions $a_{3}^{u,c}$ and $a_{5}^{u,c}$
due to perturbative QCD rescattering vanish because of the
following relations,
\begin{eqnarray}
c_{3,5}^{u,c} &=& -c_{4,6}^{u,c}/N_c = P^{u,c}_s/N_c\;,\;\;
\end{eqnarray}
where $N_c$ is the number of color.
The leading contributions to $P^i_{s}$ are given by:
 $P^i_s = ({\frac{\alpha_s}{8\pi}}) c_1 ({\frac{10}{9}} +G(m_i,\mu,q^2))$ with $i=u,c$.
 The function
$G(m,\mu,q^2)$ is given by
\begin{eqnarray}
G(m,\mu,q^2) = 4\int^1_0 x(1-x)  \mbox{ln}{m^2-x(1-x)q^2\over
\mu^2} ~\mbox{d}x \;.
\end{eqnarray}

 The rescattering via electroweak interactions are given by \cite{dattakx}
 \bea
c_{7,9}^{u,c} = P^{u,c}_e\;,\;\; c_{8,10}^{u,c} = 0
\eea
with
$P^i_e = ({\frac{\alpha_{em}}{9\pi}})
(N_c c_2+ c_1) ({\frac{10}{9}} + G(m_i,\mu,q^2))$.
These contributions are again much smaller than the dominant tree
contributions and can be neglected.

In light of the above facts we can conclude that the dominant
contributions in $X$ in Eq.~(\ref{sm}) come the tree level term
where $c_1 = 1.081$ and $c_2 = -0.190$ are the relevant Wilson
coefficients \cite{BuraseffH}. This leads to \bea X & \approx &
V_{cb} V_{cs}^* a_2= 0.17 V_{cb} V_{cs}^*\ \label{Xsm} \eea

The matrix elements in Eq.~(\ref{sm}) above can be expressed
 in terms of form factors. This can be done as follows
\cite{BSW}:
\bea \bra{V_2(k_2)} {\bar q}' \gamma_\mu b \ket{\bar{B}_s(p)}& =&
i {{2 V^{(2)}(r^2)} \over (m_B+m_{2})} \epsilon_{\mu
\nu\rho\sigma} p^\nu
k_2^\rho \varepsilon_2^{*\sigma} ~,\nn\\
\bra{V_2(k_2)} {\bar q}' \gamma_\mu\gamma_5 b \ket{B(p)}& =&
(m_B+m_{2})A_1^{(2)}(r^2) \left[\varepsilon_{2\mu}^{*}-
\frac{\varepsilon_2^{*}.r}{r^2}r_{\mu}\right]\nn\\
&& ~-A_2^{(2)}(r^2)\frac{\varepsilon_2^{*}.r}{m_B+m_{2}}
\left[(p_{\mu}+k_{2\mu})- \frac{m_B^2-m_{2}^2}{r^2}r_{\mu}\right]\nn\\
&& ~+2 im_2\frac{\varepsilon_2^{*}.r}{r^2}r_{\mu}A_0^{(2)}(r^2) ~,
\label{ffactor}
\eea
where $r=p-k_2$, and $V^{(2)}$, $A_1^{(2)}$, $A_2^{(2)}$ and
$A_0^{(2)}$ are form factors.

Using Eq. (\ref{ffactor}) in Eq. (\ref{sm}) one obtains, \bea
\bar{a}_{SM} &=& -\frac{G_F}{\sqrt{2}}m_{J/\psi}g_{J/\psi}x(m_{B_s}+m_{\phi}) A_1^{(2)}(m_{J/\psi}^2) X  \nonumber\\
\bar{b}_{SM} &=& \frac{G_F}{\sqrt{2}}2m_{J/\psi}g_{J/\psi}{m_{B_s} \over (m_{B_s}+m_{\phi})}m_{B_s} A_2^{(2)}(m_{J/\psi}^2) X \nonumber\\
\bar{c}_{SM} &=& -\frac{G_F}{\sqrt{2}}m_{J/\psi}g_{J/\psi}{m_{B_s}
\over (m_{B_s}+m_{\phi})}m_{B_s} V^{(2)}(m_{J/\psi}^2) X .\
\label{abcSM} \eea

Let us turn now to the SUSY contribution. We will consider only
the dominanat chromomagnetic operators. The gluon in these
operators can split into a charm quark pair, thereby contributing
to $ b \to s \bar{c} c$. We begin with a discussion on the matrix
elements of the chromomagnetic operators $O_g$ and
$\widetilde{O}_g$. These are given by, %
\bea
\langle J/\psi \phi| O_g |{\bar{B}_s}\rangle &=& <O_g> \nonumber\\
& =& -\frac{\alpha_s m_b}{\pi  q^2}\langle J/\psi \phi|
\left(\bar{s}_{\alpha}\gamma_{\mu}{\not {q}}(1+\gamma_5)
\frac{\lambda^A_{\alpha\beta}}{2}b_{\beta}\right)
\left(\bar{s}_{\rho}\gamma^{\mu}\frac{\lambda^A_{\rho\sigma}}{2}s_{\sigma}\right)|{\bar{B}_s}\rangle
 \nonumber\\
\langle J/\psi \phi| \widetilde{O}_g |{\bar{B}_s}\rangle &=& <\widetilde{O}_g> \nonumber\\
&=&-\frac{\alpha_s m_b}{\pi  q^2}\langle J/\psi \phi|
\left(\bar{s}_{\alpha}\gamma_{\mu}{\not {q}}(1-\gamma_5)
\frac{\lambda^A_{\alpha\beta}}{2}b_{\beta}\right)
\left(\bar{s}_{\rho}\gamma^{\mu}\frac{\lambda^A_{\rho\sigma}}{2}s_{\sigma}\right)|{\bar{B}_s}\rangle \nonumber\\
\, \label{chromo} \eea where $q^{\mu}$ is the momentum carried by
the gluon in the penguin diagram. In our case $q^{\mu}$ coincides
with the four momentum of the $J/\psi$.

After a color fierz we can write the operator $O_g$ as, \bea O_g &
= & Y_g \left[ -\frac{2}{N_c}
\left(\bar{s}_{\alpha}\gamma_{\mu}\frac{{\not{q}}}{m_b}(1+\gamma_5)
b_{\alpha}\right)
\left(\bar{s}_{\beta}\gamma^{\mu}s_{\beta}\right)+.. \right]\nonumber\\
\widetilde{O}_g & = & Y_g \left[ -\frac{2}{N_c}
\left(\bar{s}_{\alpha}\gamma_{\mu}\frac{{\not{q}}}{m_b}(1-\gamma_5)
b_{\alpha}\right)
\left(\bar{s}_{\beta}\gamma^{\mu}s_{\beta}\right)+.. \right]\nonumber\\
Y_g & = & -\frac{\alpha_s m_b^2}{4\pi  m_{J/\psi}^2}. \nonumber\
\label{dipole} \eea In the above we have only retained terms that
contribute to the decay $\bar{B}_s(p) \to J/\psi(k_1,
\varepsilon_1) \phi(k_2,\varepsilon_2)$ In factorization, after
using equation of motion, we can write the matrix element of $O_g$
as, \bea
<O_g> & = & T_1+T_2+T_3 \nonumber\\
T_1 & = & C_gY_g \left[ -\frac{2}{N_c}L_{J/\psi} \right] \nonumber\\
L_{J/\psi} & = & m_{J/\psi} g_{J/\psi} \varepsilon_1^{*\mu}
\bra{\phi} \bar{s} \gamma_{\mu} ( 1-\gamma_5 ) b \ket{\bar{B}_s},\nonumber\\
T_2 & = & C_gY_g\frac{m_s}{m_b} \left[ -\frac{2}{N_c}R_{J/\psi} \right] \nonumber\\
R_{J/\psi} & = & m_{J/\psi} g_{J/\psi} \varepsilon_1^{*\mu}
\bra{\phi} \bar{s} \gamma_{\mu} ( 1+\gamma_5 ) b \ket{\bar{B}_s},\nonumber\\
T_3 & = &C_gY_g\frac{2 \varepsilon_1^* \cdot k_2}{m_b} \left[ \frac{2}{N_c}S_{J/\psi} \right] \nonumber\\
S_{J/\psi} & = & m_{J/\psi} g_{J/\psi} \bra{\phi} \bar{s}  (
1+\gamma_5 ) b \ket{\bar{B}_s}.\ \label{susy1} \eea In the above
$m_{s,b}$ are the strange and the bottom quark masses.

In the above equation it is clear that $T_2$ is suppressed
relative to $T_1$ by ${ m_s \over m_b}$ and we will neglect it.
From the structure of the polarization vectors in
Eq.~(\ref{pol_phi}), it is also clear that the $\pm$ polizations
do not contribute to $T_3$. Hence for the $\pm$ polarizations we
can obtain a clear prediction for $r^{\lambda}_A$ defined below
Eq.~(\ref{newbetas}), as the form factors and other hadronic
quantities cancel in the ratio.

{}For the matrix element of the operator $\widetilde{O}_g$,
focussing  only on the transverse amplitues we can write, \bea
<\widetilde{O}_g> & =  & Y_g \left[ -\frac{2}{N_c}R_{J/\psi} \right] \nonumber\\
R_{J/\psi} & = & m_{J/\psi} g_{J/\psi} \varepsilon_1^{*\mu}
\bra{\phi} \bar{s} \gamma_{\mu} ( 1+\gamma_5 ) b \ket{\bar{B}_s},\
\label{susy2} \eea

Hence again focussing only on the transverse amplitudes we can
write, using Eq.~(\ref{ffactor}) in Eq.~(\ref{susy1}) and
Eq.~(\ref{susy2}), \bea
\bar{a}_{susy} &=& -\frac{G_F}{\sqrt{2}}m_{J/\psi}g_{J/\psi}(m_{B_s}+m_{\phi}) A_1^{(2)}(m_{J/\psi}^2) (Y- \tilde{Y})  \nonumber\\
\bar{c}_{susy} &=& -\frac{G_F}{\sqrt{2}}m_{J/\psi}g_{J/\psi}{m_{B_s} \over (m_{B_s}+m_{\phi})}m_{B_s} V^{(2)}(m_{J/\psi}^2) (Y+ \tilde{Y}) \nonumber\\
Y & = &\frac{\sqrt{2}C_g}{G_F}Y_g \left[ -\frac{2}{N_c}\right] \nonumber\\
\widetilde{Y} & = &\frac{\sqrt{2}\widetilde{C_g}}{G_F}Y_g \left[ -\frac{2}{N_c}\right] \nonumber\\
Y_g & = & -\frac{\alpha_s m_b^2}{4\pi  m_{J/\psi}^2}\
\label{abcSUSY}
\eea
Combining the SM and SUSY contributions we can now compute,
\bea
r_A^{\|} & = & \vert A_{{\rm NP}}^{\|}/A_{{\rm SM}}^{\|} \vert
=\vert\frac{(Y- \widetilde{Y})}{X}\vert\nonumber\\
r_A^{\perp} & = & \vert A_{{\rm NP}}^{\perp}/A_{{\rm SM}}^{\perp} \vert
=\vert\frac{(Y+ \widetilde{Y})}{X}\vert\
\label{ratio}
\eea
Using the values of $V_{cb}$ and $V_{cs}$ from Ref~\cite {pdg} we obtain
$X\approx 0.0069$. Futhermore with $m_{\widetilde g}=m_{\widetilde q}=500$ GeV, $m_b(m_b)=4.5$ GeV, we obtain,
\bea
Y & \approx &= 2.1315 (\delta^d_{LR})_{23} \left[\frac{-2}{N_c}Y_g\right] =
0.0477(\delta^d_{LR})_{23}\nonumber\\
\widetilde{Y} & \approx &= 2.1315 (\delta^d_{RL})_{23} \left[\frac{-2}{N_c}Y_g\right] =0.0477(\delta^d_{RL})_{23}\
\label{ynum}
\eea
We can then write, using Eq.~\ref{ratio},
\bea
r_A^{\|} & \approx &
 0.07
\frac{\sqrt{(\vert\lr\vert)^2+ (\vert \rl \vert)^2 - 2 \vert\lr\vert\vert \rl \vert \cos (\theta_{LR}-\theta_{RL})}}{0.01}\nonumber\\
r_A^{\perp} & \approx &  0.07 \frac{\sqrt{(\vert\lr\vert)^2+
(\vert \rl \vert)^2 + 2 \vert\lr\vert\vert \rl \vert \cos
(\theta_{LR}-\theta_{RL})}}{0.01},\ \label{rationum} \eea where
$\theta_{LR}$ and $\theta_{RL}$ are the phases of $\lr$ and $\rl$.
We will set $\vert \lr \vert= \vert \rl \vert =0.01$ and we can
then now consider the following cases:\\

{\bf {Case a}} $\lr=\rl$. In this case we obtain,
\bea %
S_{J/\psi \phi}^{\|} &=& \sin(2 \beta_s^{SM} + 2 \theta_s) \nonumber\\
S_{J/\psi \phi}^{\perp} &=& \sin(2 \beta_s^{SM} + 2 \theta_s) + 0.28 \cos(2 \beta_s^{SM}
+ 2 \theta_s) \sin \theta^{\perp}_{{\rm NP}} \cos \delta^{\perp}.\
\label{newbetasnuma} %
\eea%
If we neglect the contribution from mixing then $S_{J/\psi
\phi}^{\perp}$ can reach a value of upto 0.3 for $\sin
\theta^{\perp}_{{\rm NP}} \sim 1$ and $\cos \delta^{\perp}\sim
1$.\\

{\bf {Case b}} $\lr=-\rl$. In this case we obtain,
\bea %
S_{J/\psi \phi}^{\|} &=& \sin(2 \beta_s^{SM} + 2 \theta_s) + 0.28 \cos(2 \beta_s^{SM}
+ 2 \theta_s) \sin \theta^{\|}_{{\rm NP}} \cos \delta^{\|}\nonumber\\
S_{J/\psi \phi}^{\perp} &=& \sin(2 \beta_s^{SM} + 2 \theta_s). \
\label{newbetasnumb} %
\eea%
Again, if we neglect the contribution from mixing then $S_{J/\psi
\phi}^{\|}$ can reach a value of upto 0.3 for $\sin
\theta^{\|}_{{\rm NP}} \sim 1$ and $\cos \delta^{\|}\sim 1$.
Finally, we can consider the case where either $\lr$ or $\rl$ is
zero. For the case $\lr \ne 0, \rl = 0$ we obtain,
\bea %
S_{J/\psi \phi}^{\|} &=& \sin(2 \beta_s^{SM} + 2 \theta_s) + 0.14 \cos(2 \beta_s^{SM}
+ 2 \theta_s) \sin \theta^{\|}_{{\rm NP}} \cos \delta^{\|}\nonumber\\
S_{J/\psi \phi}^{\perp} &=&
\sin(2 \beta_s^{SM} + 2 \theta_s) + 0.14 \cos(2 \beta_s^{SM}
+ 2 \theta_s) \sin \theta^{\perp}_{{\rm NP}} \cos \delta^{\perp}\
\label{newbetasnum1} %
\eea%
For the case $\lr=0, \rl \ne 0$ we obtain,
\bea %
S_{J/\psi \phi}^{\|} &=& \sin(2 \beta_s^{SM} + 2 \theta_s) - 0.14 \cos(2 \beta_s^{SM}
+ 2 \theta_s) \sin \theta^{\|}_{{\rm NP}} \cos \delta^{\|}\nonumber\\
S_{J/\psi \phi}^{\perp} &=&
\sin(2 \beta_s^{SM} + 2 \theta_s) + 0.14 \cos(2 \beta_s^{SM}
+ 2 \theta_s) \sin \theta^{\perp}_{{\rm NP}} \cos \delta^{\perp}\
\label{newbetasnum2} %
\eea%

Now one may wonder how NP in $ b \to s \bar{c} c$ transitions
affect CP measurements in the $B_d$ system. Let us first consider
the indirect CP asymmetry in the golden mode $ B_d \to J/\psi
K_s$. Note this is a vector-pseudoscalar decay and so the strong
phases involved here can be quite different from the ones involved
in vector-vector decays. In other words NP effects in different
final states can be very different. More interestingly, it can be
easily checked that for case b in Eq.~(\ref{newbetasnumb}) the
contribution to the indirect asymmetry in $ B_d \to J/\psi K_s$
cancels. However, the indirect CP asymmetry in the vector-vector
mode does not cancel for all polarization states. In other words
the range of NP effects obtained in the decay $B_s \to J/\psi
\phi$ are consistent with $\sin{2 \beta}$ measurements in $B_d \to
J/\psi K_s$ \cite{hfag,utfit,ckmfitter} for the various reasons
discussed above.

The decay $B_d \to J/\psi K*$ is related to $B^0_s \to J/\psi
\phi$ by $SU(3)$ flavor symmetry. Hence we should potentially see
NP effects in $B_d \to J/\psi K*$ , up to $SU(3)$ breaking
effects. The CP measurements in this decay are not yet precise
\cite{hfag} and hence this decay also is an ideal place to look
for new physics effects in the decay amplitude.

\subsection{\large{\bf Summary}}
In summary, we have analyzed the SUSY contribution to $B^0_s
-\bar{B}^0_s$ mixing in light of recent experimental measurement
of the mixing phase. We showed that the experimental limits of the
mass difference $\Delta M_{B_s}$ and the mercury EDM constrain
significantly the SUSY contribution to $B^0_s -\bar{B}^0_s$
mixing, so that $\sin 2 \beta_s \lsim 0.1$. We then studied the
 the one loop SUSY contribution to $B^0_s \to J/\psi
\phi$ decay and found that new physics contribution to the decay
amplitude can lead to significant indirect CP asymmetries which
are in general different for different polarization states.

\section*{Acknowledgments} We would like to thank  A. Masiero for
fruitful discussions. The work of S.K. was partially supported by
the ICTP grant Proj-30 and the Egyptian Academy for Scientific
Research and Technology.

%%%%%%%%%%%%%%%%%%%%%%%%%%%%%%%%%%%%%%%%%%%

\end{document}